\input psfig.sty
\documentclass[referee,letterpaper]{aa}
\thesaurus{07.16.1}
\title{Measuring $\Omega_{\rm m}$ using clusters evolution}
\author{A. Del Popolo\inst{1,2,3}
}
\offprints{A. Del Popolo - {\it E-mail adelpop@unibg.it}}
\institute{$^1$ Dipartimento di Matematica, Universit\`{a} Statale
di Bergamo,
via dei Caniana, 2 - I 24129 Bergamo, ITALY \\
$^2$ Feza G\"ursey Institute, P.O. Box 6 \c Cengelk\"oy, Istanbul, Turkey \\
$^3$ Bo$\breve{g}azi$\c{c}i University, Physics Department,
80815 Bebek, Istanbul, Turkey\\
}
\titlerunning{Measuring $\Omega_{\rm m}$ using clusters evolution}
\authorrunning{A. Del Popolo}
\date{}

\begin{document}
\maketitle

\begin{abstract}

The constraints obtained
by several authors (Eke et al. 1998; Henry 2000)
on the estimated values of $\Omega_{\rm m}$, $n$ and $\sigma_8$ are revisited in the light of recent theoretical developments: 1) new theoretical mass functions; 2) a more accurate mass-temperature relation, also determined for arbitrary $\Omega_{\rm m}$ and $\Omega_{\rm \Lambda}$.
I re-derive the X-ray Temperature Function (XTF), similarly to Henry (2000) and Eke et al. (1999), re-obtaining the constraints on $\Omega_{\rm m}$, $n$, $\sigma_8$.
The result shows that changes in the mass function and M-T relation produces an increase in $\Omega_{\rm m}$ of $ \simeq 20\%$ and similar results in $\sigma_8$ and $n$.
\end{abstract}

\begin{keywords}
cosmology: theory - large scale structure of Universe - galaxies:
formation
\end{keywords}

\section{Introduction}

Galaxy clusters represents the virialization stage of exceptionally high peaks of initial density perturbations
on comoving scales of $\simeq 10 h^{-1} {\rm Mpc}$, and as such they provide useful cosmological probes. The evolution
in the abundance of clusters is strongly dependent on the cosmological density parameter, $\Omega_{\rm m}$ (Evrard 1989; Oukbir \& Blanchard 1992; Eke et al. 1996; Donahue et al. 1998; Borgani et al. 1999).
X-ray observations provide a very efficient method to identify distant clusters down to a given X-ray flux limit, and hence within a known survey volume for each luminosity, $L_{\rm X}$. For this reason, most studies using clusters as cosmological probes are based on X-ray samples.
It is well known that clusters are strong X-ray emitters and so cluster evolution can be inferred from the study of X-ray properties of distant clusters. The amount of observational data concerning high-redshift cluster properties has increased in the past years. {\it EMSS} (Henry et al. 1992; Gioia \& Luppino 1994), {\it ASCA} (Donahue 1996; Henry 1997) and {\it ROSAT} (Ebeling et al. 1997; Rosati et al. 1998)
In addition, galaxy velocity dispersions for a well-defined sample of high-redshift clusters (Carlberg et al. 1996) are provided by the {\it CNOC} survey. The Press \& Schechter (1974) (hereafter PS) formalism has been heavily used to model the cluster population. The combination of the PS mass function and the X-ray cluster catalogs represents a unique opportunity to constraint cosmological parameters, (e.g. the mass density parameter, $\Omega_{\rm m}$).
Although the analytical framework of the PS model has been greatly
refined and extended (e.g., Lacey \& Cole 1993), it is well known that the PS mass function,
while qualitatively correct, disagrees with the results of
N-body simulations.
The quoted discrepancy is not surprising since the PS model, as any other analytical model,
should make several assumptions to get simple analytical predictions.

There are different methods to trace the evolution of the cluster number density: \\
a) The X-ray temperature function (XTF) has been presented for local (e.g. Henry \& Arnaud 1991) and distant clusters (Eke et al. 1998; Henry 2000). The mild evolution of the XTF has been interpreted as a strong indication for a low density universe ($0.2<\Omega_{\rm m}<0.6$). \\
b) The evolution of the X-ray luminosity function (XLF).

The results for $\Omega_{\rm m}$ obtained span the entire range of acceptable solutions: $0.2 \leq \Omega_{\rm m} \leq 1$ (see Reichart et al. 1999).
The reasons leading to the quoted discrepancies has been studied in several papers (Eke et al. 1998; Reichart et al. 1999; Donahue \& Voit 1999; Borgani et al. 2001).

Although the quoted uncertainties has been so far of minor importance with respect to the paucity of observational data, a breakthrough is needed in the quality of the theoretical framework if high-redshift clusters are to take part in the high-precision-era of observational cosmology.


These reasons lead me to re-calculate the constraints on $\Omega_{\rm m}$, $n$ and $\sigma_8$, using the XTF.
In Sect. ~2, I re-calculate the XTF, as done by Henry (2000) and Eke et al. (1998) and obtained
constraints for $\Omega_{\rm m}$ and $n$ and $\sigma_8$.
Sect. ~3 is devoted to results and to conclusions.

\section{Constraints to cosmological parameters from the XTF}

The mass function (MF) is a critical ingredient in putting strong constraints on
cosmological parameters (e.g., $\Omega_{\rm m}$).
Observationally the local mass function has been derived from
measuring masses of individual clusters from galaxy velocity dispersions or
other optical properties by Bahcall and Cen (1993), Biviano et al. (1993), and
Girardi et al. (1998). However, the estimated virial masses for individual clusters
depend rather strongly on model assumptions.
As argued by Evrard et al. (1996) on the basis of hydrodynamical
N-body simulations, cluster masses may be presently more accurately
determined from a temperature measurement
and a mass-temperature relation determined from detailed observations
or numerical modeling.
Thus alternatively, as a well-defined observational quantity,
the X-ray temperature function (XTF) has been measured,
which can be converted to the MF by means of the mass-temperature relation.
\begin{figure}[tbp]
\psfig{figure=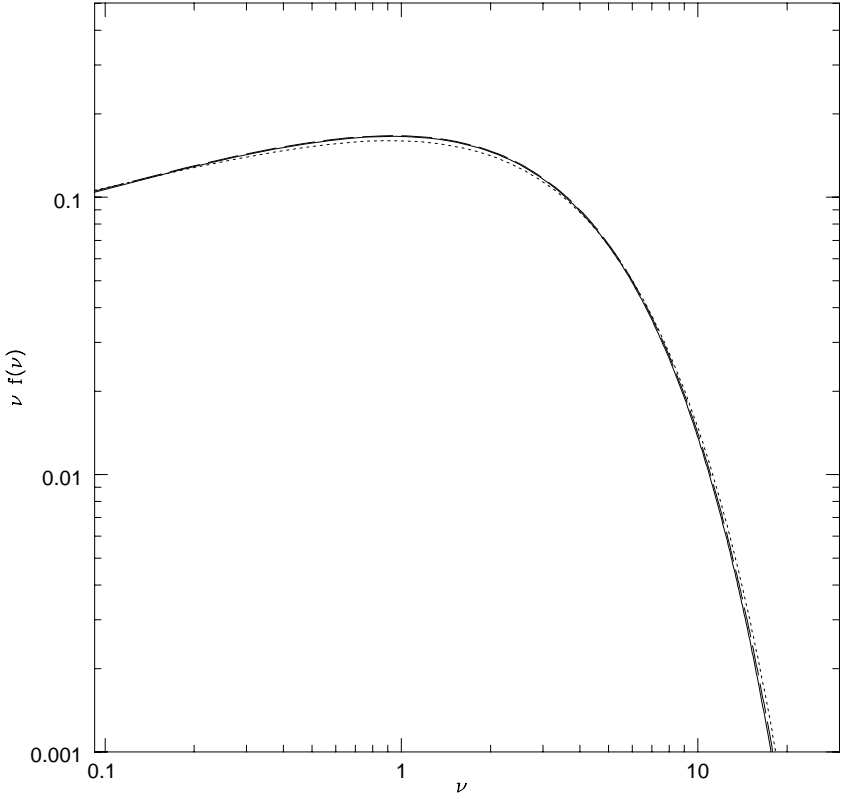,width=12cm} 
\caption{(a) Comparison of various mass functions. The dotted line represents Sheth \& Tormen (2002) prediction, the solid line that of Jenkins et al. (2001) and the dashed line that of Del Popolo (2000b).}
\end{figure}

The cluster temperature function is defined as:
\begin{equation}
N(T,z)=N(M,z) \frac{d M}{dT}
\label{eq:temppp}
\end{equation}
While the mass function, $N(M,z)$, gives the mass and distribution of a population of
evolving clusters, the Jacobian $\frac{d M}{dT}$ describes the physical properties of the single cluster.

Comparison of the predictions of the PS theory with the SCDM and OCDM cosmologies, performed by
Tozzi \& Governato (1998) and Governato et al. (1999), have shown discrepancies between PS predictions and
N-body simulations, increasing with increasing $z$.

\begin{figure}[tbp]
\centerline{\hbox{(a)
\psfig{figure=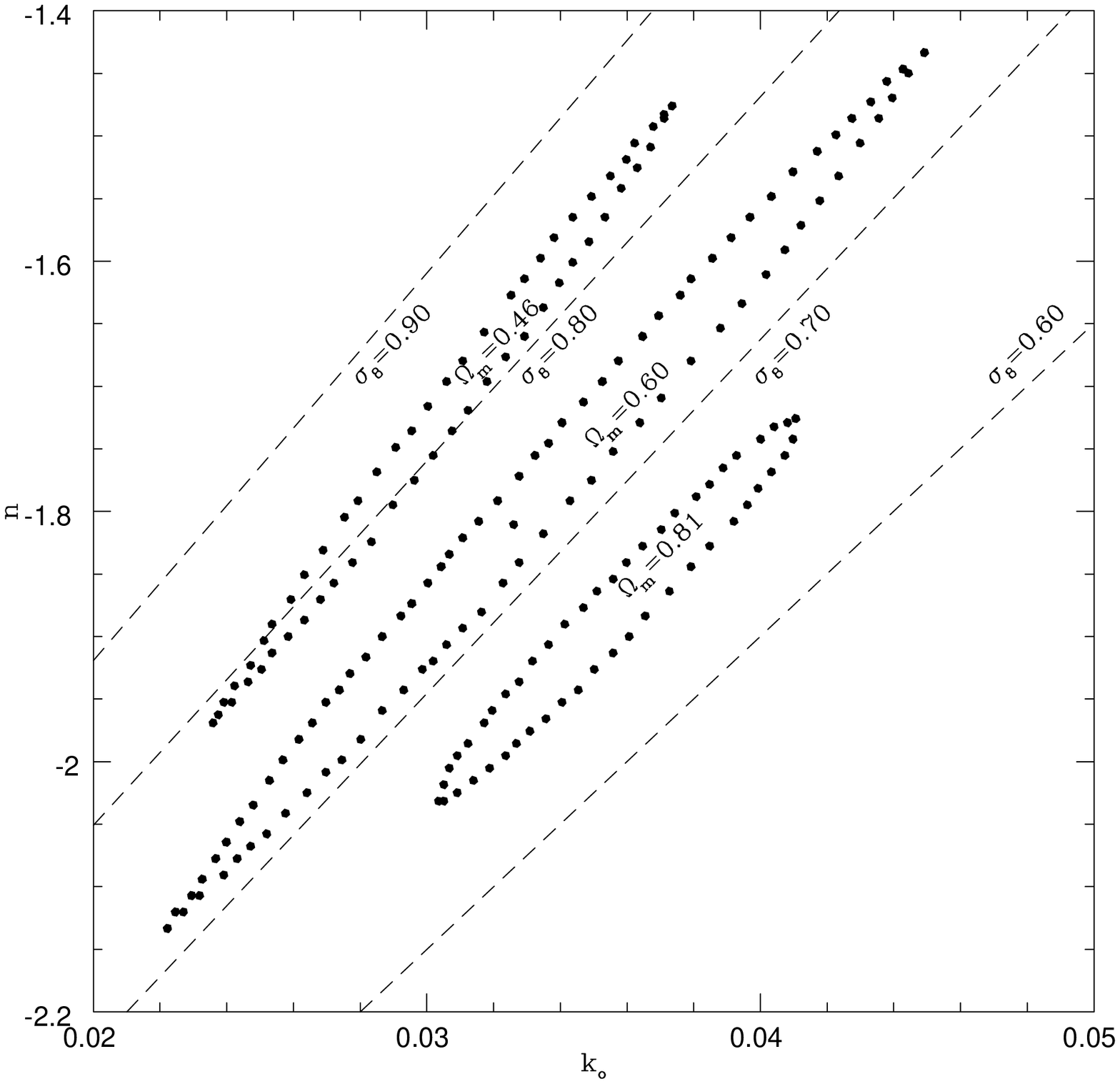,width=6cm} (b)
\hspace{1cm}
\psfig{figure=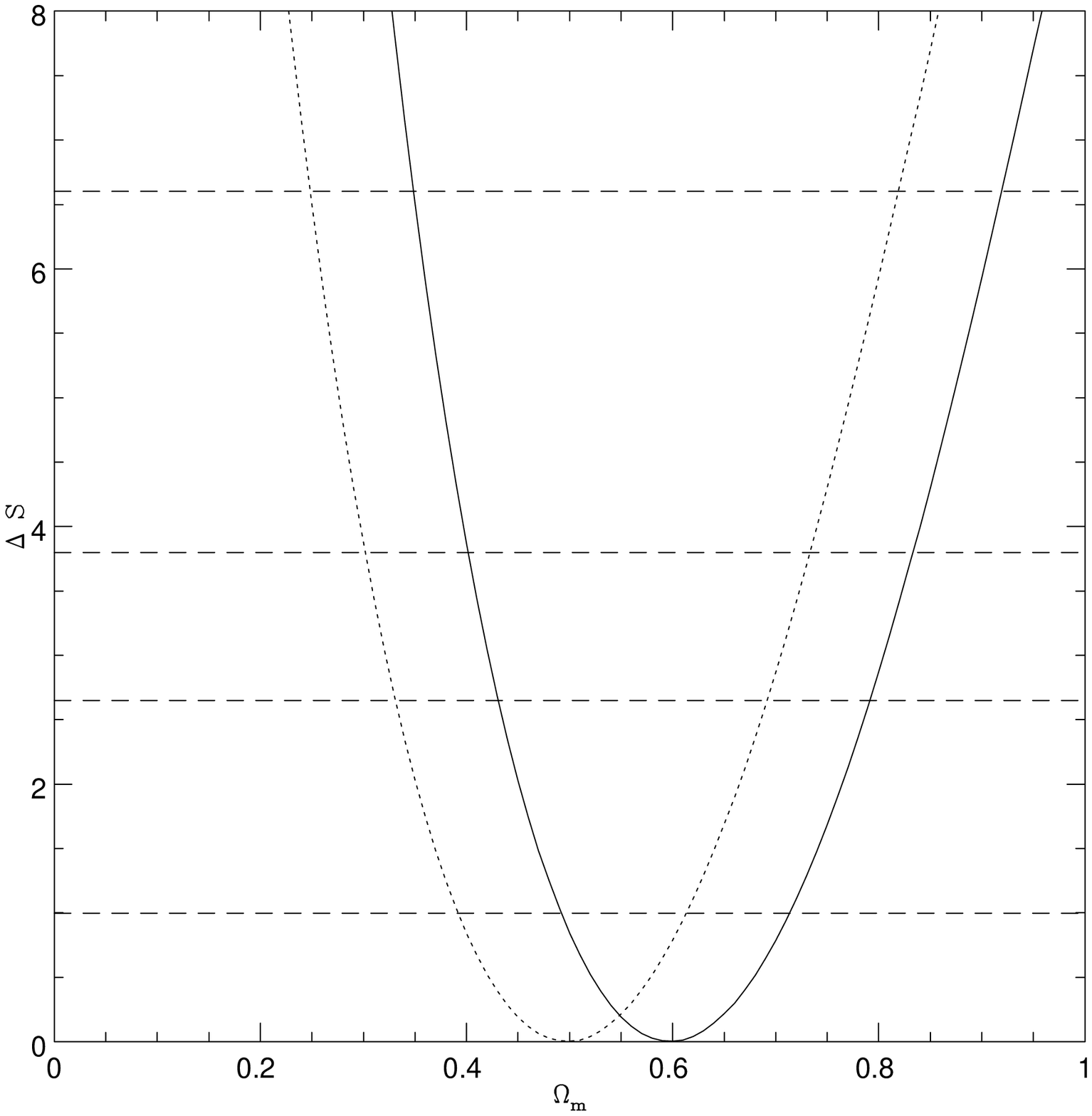,width=6cm}
}}
\centerline{\hbox{(c)
\psfig{figure=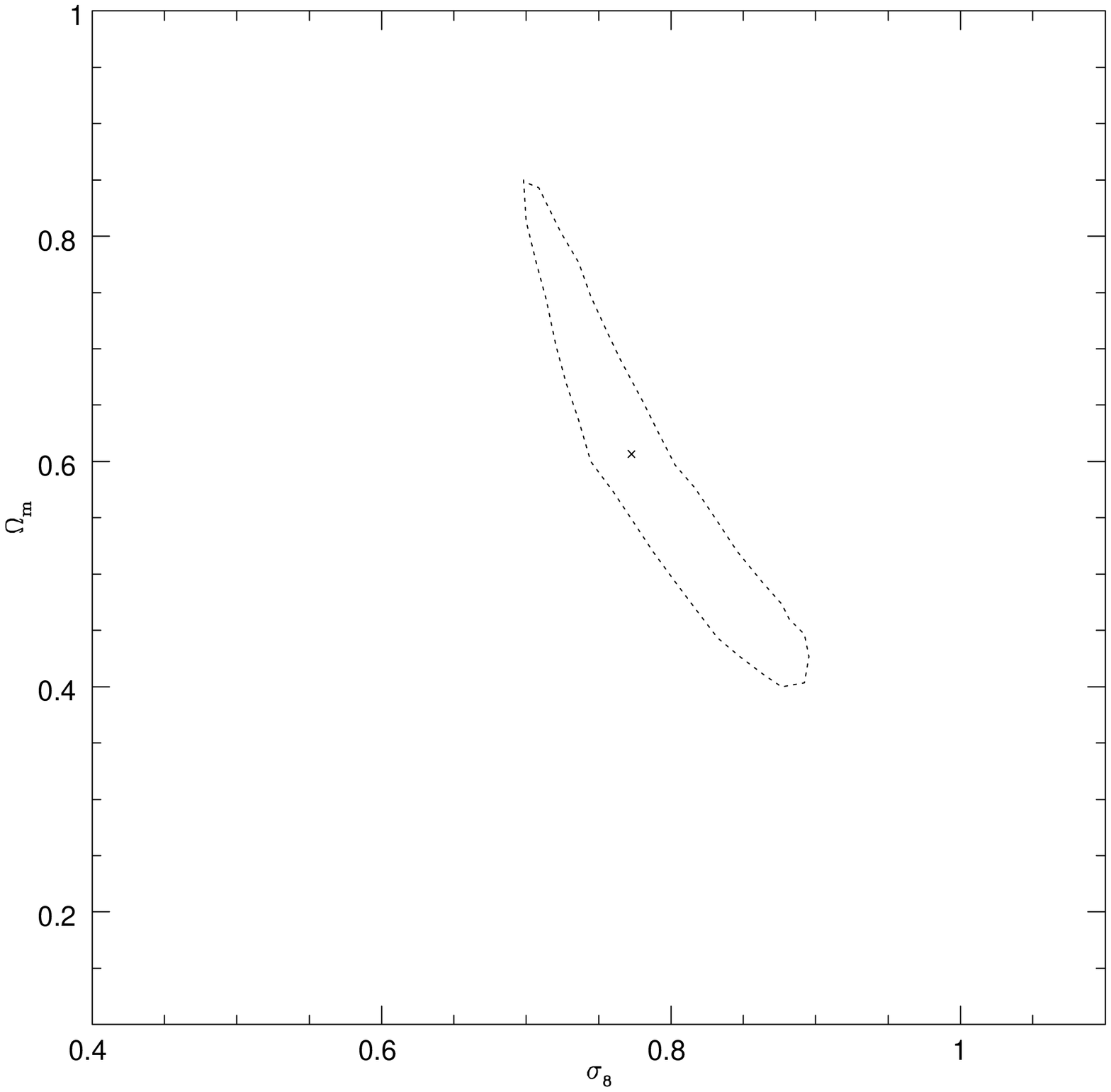,width=6cm} (d)
\hspace{1cm}
\psfig{figure=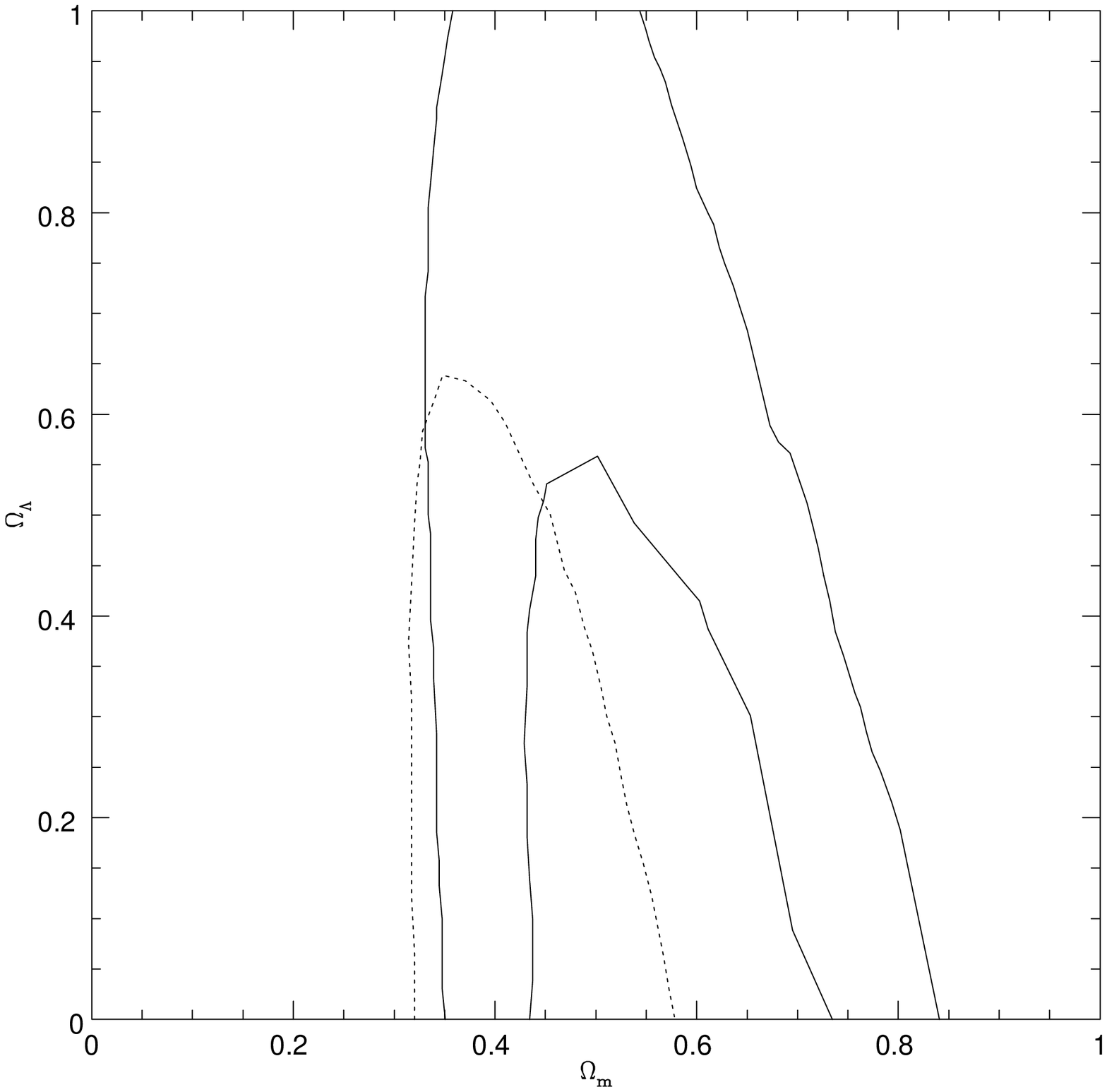,width=6cm}
}}
\caption{(a) The $68\%$ confidence contours for the parameters $n$, $k_{\rm o}$ and $\Omega_{\rm m}$ for the open model. The dashed lines are lines of constant $\sigma_8$. (b)
$\Delta$(likelihood)
for
the parameter $\Omega_{\rm m}$. The solid line is obtained from the model of this paper while the dotted line is that calculated by Henry (2000). The dashed lines represent various confidence levels (65\%, 90\%, 95\%, 99\%). 
(c)The $68\%$ confidence contours for the parameters $\sigma_8$, and $\Omega_{\rm m}$ for the open model (see also Henry (2000), Fig. 9).
(d)Constraints on $\Omega_{\Lambda}$ and $\Omega_{\rm m}$ obtained using the same 25 clusters used in Henry (2000), for the local sample, while the high redshift sample is constituted from all the EMSS clusters with $z>0.3$ and RX J0152.7-1357
(see Henry 2002). The solid lines are the 1 and 2 $\sigma$ contours obtained using the mass function and the M-T relation of this paper, while the dashed line is the 1 $\sigma$ contour obtained using the PS mass function and the M-T relation of Pierpaoli et al. 2001.}
\end{figure}

In the following, I'll use the mass function modified as
described in Del Popolo (2000b)
and an improved form of the M-T relation in order to calculate the mass function (Del Popolo 2000a).
These are respectively given by:
\begin{equation}
n(m,z)\simeq 1.21 \frac{\overline{\rho}}{m^{2}}\frac{d\log (\nu )}{d\log m}
\left( 1+\frac{0.06}{\left( a\nu \right) ^{0.585}}\right) \sqrt{\frac{a\nu }{2\pi }}\exp{\{-a\nu \left[ 1+\frac{0.57}{\left( a\nu \right) ^{0.585}}\right] ^{2}/2\}}
\label{eq:nmm}
\end{equation}
where $a=0.707$.
and
\begin{equation}
kT \simeq 8 keV \left(\frac{M^{\frac 23}}{10^{15}h^{-1} M_{\odot}}\right)
\frac{
\left[
\frac{1}{m_1}+\left( \frac{t_\Omega }t\right) ^{\frac 23}
+\frac{K(m_1,x)}{M^{8/3}}
\right]
}
{
\left[
\frac{1}{m_1}+\left( \frac{t_\Omega }{t_{0}}\right) ^{\frac 23}
 +\frac{K_0(m_1,x)}{M_{0}^{8/3}}
\right]
}
\label{eq:kTT1}
\end{equation}
(see Del Popolo 2000a for a derivation),
where $M_{0} \simeq 5 \times 10^{14} h^{-1} M_{\odot}$,
$t_\Omega =\frac{\pi \Omega _{\rm m}}{H_o\left( 1-\Omega _{\rm m}-\Omega _\Lambda \right) ^{\frac 32}}$,
$m_1=5/(n+3)$ (being $n$ the spectral index), and:
\begin{eqnarray}
K(m_1,x)&=&F x \left (m_1-1\right ) {\it LerchPhi}
(x,1,3m_1/5+1)-
\nonumber \\
& &
F \left (m_1-1\right ){\it LerchPhi}(x,1,3m_1/5)
\end{eqnarray}
where $F$ is defined in Del Popolo (2000a, b) and the
{\it LerchPhi} function is defined as follows:
\begin{equation}
LerchPhi(z,a,v)=\sum_{n=0}^{\infty} \frac{z^n}{(v+n)^a}
\end{equation}
where $x=1+(\frac{t_{\Omega}}{t})^{2/3}$ which is connected
to mass by $M=M_{\rm 0} x^{-3 m/5}$ (V2000),
and where $K_0(m_1,x)$ indicates that $K(m_1,x)$ must be calculated assuming $t=t_0$.
Since Eq. (\ref{eq:nmm}) and Eq. (\ref{eq:kTT1}) where obtained in previous papers, Del Popolo 2002a,b, and 
since there is a wide description of the underlying assumptions, I refer the reader to those papers for 
details.

Before going on, I want to discuss about the use of Eq. (\ref{eq:nmm}).
Jenkins et al. (2001), obtained a mass function, which is regarded as perhaps the
most accurate model to date, from the Hubble volume simulations. 
%
%
Jenkins et al. (2001) showed that, although the mass functions in their 
simulations scaled in accordance with the excursion set prediction, 
Sheth \& Tormen 2002 (Eq. (2)) slightly overestimated the unconditional mass functions 
in their simulations. It interesting to note that as shown by Pierpaoli et al. (2001), using Sheth \& Tormen (2002) or Jenkins et al. (2001, 2003) mass function, rather than PS, the value of parameters like $\sigma_8$ are changed by a small amount (4-8 \%).
Sheth \& Tormen (2002) also showed that 
changing the parameter $a$ in their Eq. (2) from 0.707 to 0.75 reduces the 
discrepancy between it and the simulations substantially. 
In Del Popolo (2002a), I showed that Eq. (\ref{eq:nmm}) of this paper, in agreement with Jenkins et al. (2001), predicts smaller values
of the mass function expecially for high $\nu$, with respect with Sheth \& Tormen's predictions.
In other terms, Eq. (\ref{eq:nmm}) of this paper is in very good agreement with Jenkins et al. (2001). In Fig. 1 I plot a comparison of the various mass functions: the dotted line represents Sheth \& Tormen (2002) prediction, the solid line that of Jenkins et al. (2001) and the dashed line that of Del Popolo (2000b). As shown, Jenkins et al. (2001) mass function is almost indistinguishable from that used in this paper. 
For what concerns small masses, our formula, in agreement with Sheth \& Tormen 2002, differs dramatically from the one Jenkins et al. (2001), propose. 
Simulations currently available do not probe the regime where $\nu \leq 0.3$ or so (the Jeans mass is at 
about $\nu \leq 0.03$ (Sheth \& Tormen 2002)). 
New simulations are needed to address which low mass behavior is correct.
In other terms, the mass function obtained in this paper is in very good agreement with Jenkins et al. (2001) in 
the regime probed by simulations. 
Moreover, the constraints obtained in almost all the papers in literature used the PS mass function except a few papers (e.g. Borgani et al. 2001; Henry 2002) and the M-T relation the usual one obtained from the virial theorem. In other words,
this paper introduces noteworthy improvements on the previous calculations in literature.
The mass entering in Eq. (\ref{eq:nmm}) should be interpreted as the mass contained inside a radius, $r_{180}$ encompassing a mean overdensity $\rho= 180 \overline{\rho}$. However, scaling relations connecting mass X-ray observable quantities may provide the mass at different values of $\rho/\overline{\rho}$. In this case we follow White (2001) and rescale the masses assuming an NFW (Navarro, Frenk \& White 1996) profile for the dark matter halo with a concentration $c=5$ appropriate for rich clusters (see also Pierpaoli et al. 2001, 2003; Schuecker et al. 2002 for more details).

Similarly, it necessary to give arguments to use Eq. (\ref{eq:kTT1}) instead of
the new mass/X-ray temperature relations obtained from simulations or
Chandra data within the last year (see, e.g., Pierpaoli et al. 2001, 2003
for a reference). As shown in Del Popolo (2002a), Eq. (\ref{eq:kTT1}) reduces to a similar equation to that 
used in Pierpaoli et al. (2001) (Eq. 13), in the early-time, namely:
\begin{equation}
M \propto T^{3/2} \rho^{-1/2} \Delta_{\rm c}^{-1/2} \propto T^{3/2} (\Delta_{\rm c} E^2)^{-1/2}
\label{eq:pierp}
\end{equation}
where $E(z)^2=\Omega_{\rm M} (1+z)^3+\Omega_{\Lambda}$, 
and the term depending on $\Omega_{\Lambda}$ in Eq. (13) of Pierpaoli et al. (2001) (which is a correction to the 
virial relation arising from the additional $r^2$ potential in the presence of $\Lambda$) is neglected since it produces only a small correction (see Pierpaoli et al. 2001). 
I also want to add that Eq. (13) of Pierpaoli et al. (2001) or Eq. (4) of Pierpaoli et al. (2003), 
comes from rather simplistic arguments (dimensional analysis and an assumption that clusters are self-similar) 
and is a good approximation to both observations and simulations
but this last are 
sufficiently computationally demanding that they cannot explore parameter space efficiently and so it is necessary to determine coefficients by means of simulations, while scalings are taken from simple theoretical models (Pierpaoli et al. 2001). 
Eq. (13) of Pierpaoli et al. (2001), is valid for systems hotter than about 3 keV.
We know that recent studies have shown that the self-similarity in the M-T relation seems to break at 
some keV (Nevalanien et al. 2000; Xu, Jin \& Wu 2001): Afshordi \& Cen (2001) has shown that non-sphericity introduces an asymmetric, mass dependent, scatter for the M-T relation altering its slope at the low mass end ($T \sim 3$ keV). These effects are taken into account by mine M-T relation, which as previously told gives same results of that of
Pierpaoli et al. (2001, 2003) in the range of energy in which this is valid.
For what concerns the definition of masses in the M-T relation, the mass is defined, according to Del Popolo (2000a), Voit \& Donahue 1998 and Voit 2000, as $M_{200}$. In order to combine the mass function and M-T relation, I performed a mass transformation, as above reported (see also Pierpaoli et al. 2001, 2003; Schuecker et al. 2002 for more details).
  
Introducing Eq. (\ref{eq:nmm}) into Eq. (\ref{eq:temppp}) and using the M-T relation, in the peculiar case that
the variance is given by:
\begin{equation}
\sigma=C (\frac{M}{10^{15} M_{\odot}})^{-(3+n)/6}=C m^{-(3+n)/6}
\end{equation}
where $n$ is the spectral index.

\begin{figure}[tbp]
\centerline{\hbox{(a)
\psfig{figure=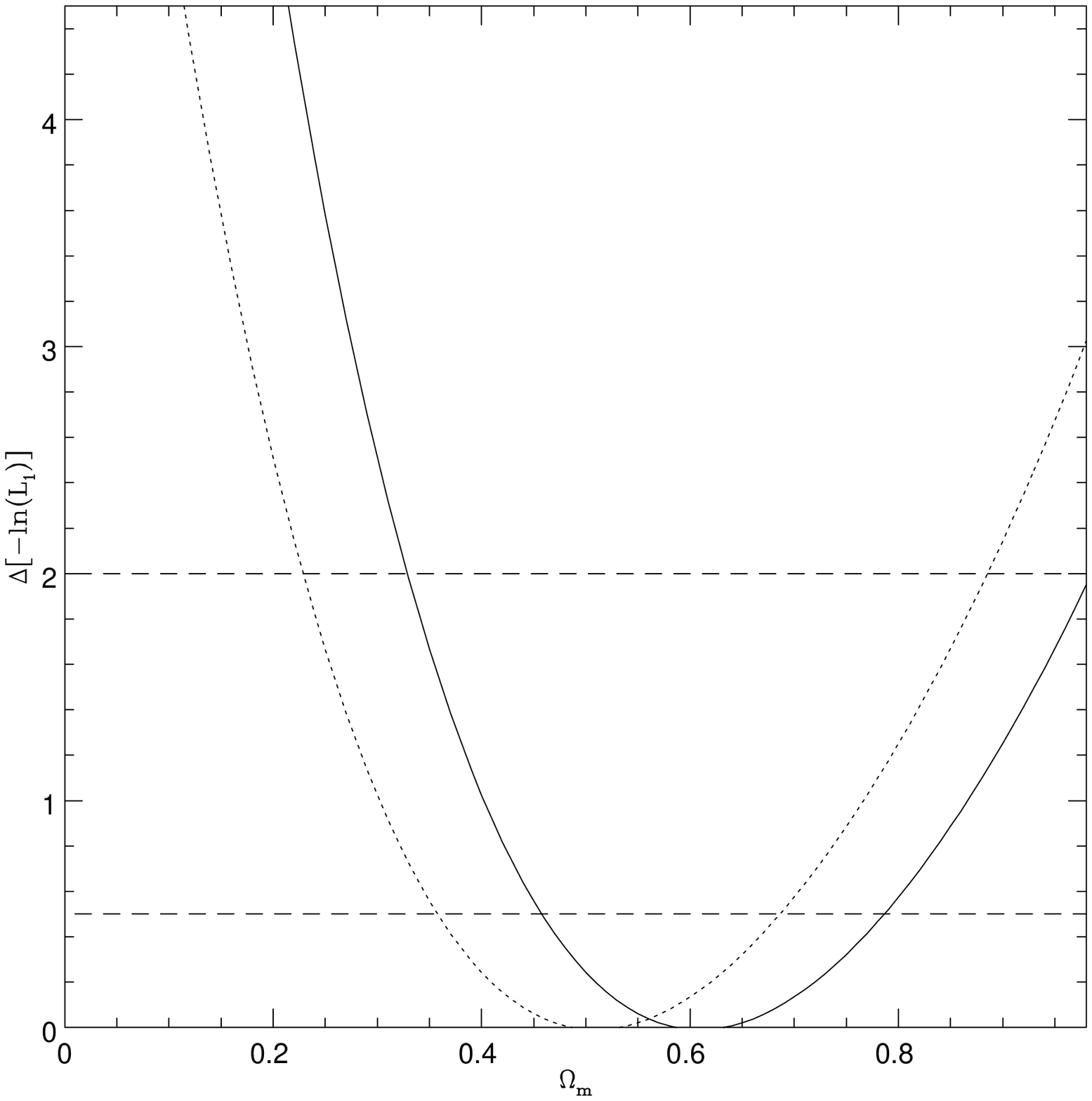,width=6cm} (b)
\hspace{1cm}
\psfig{figure=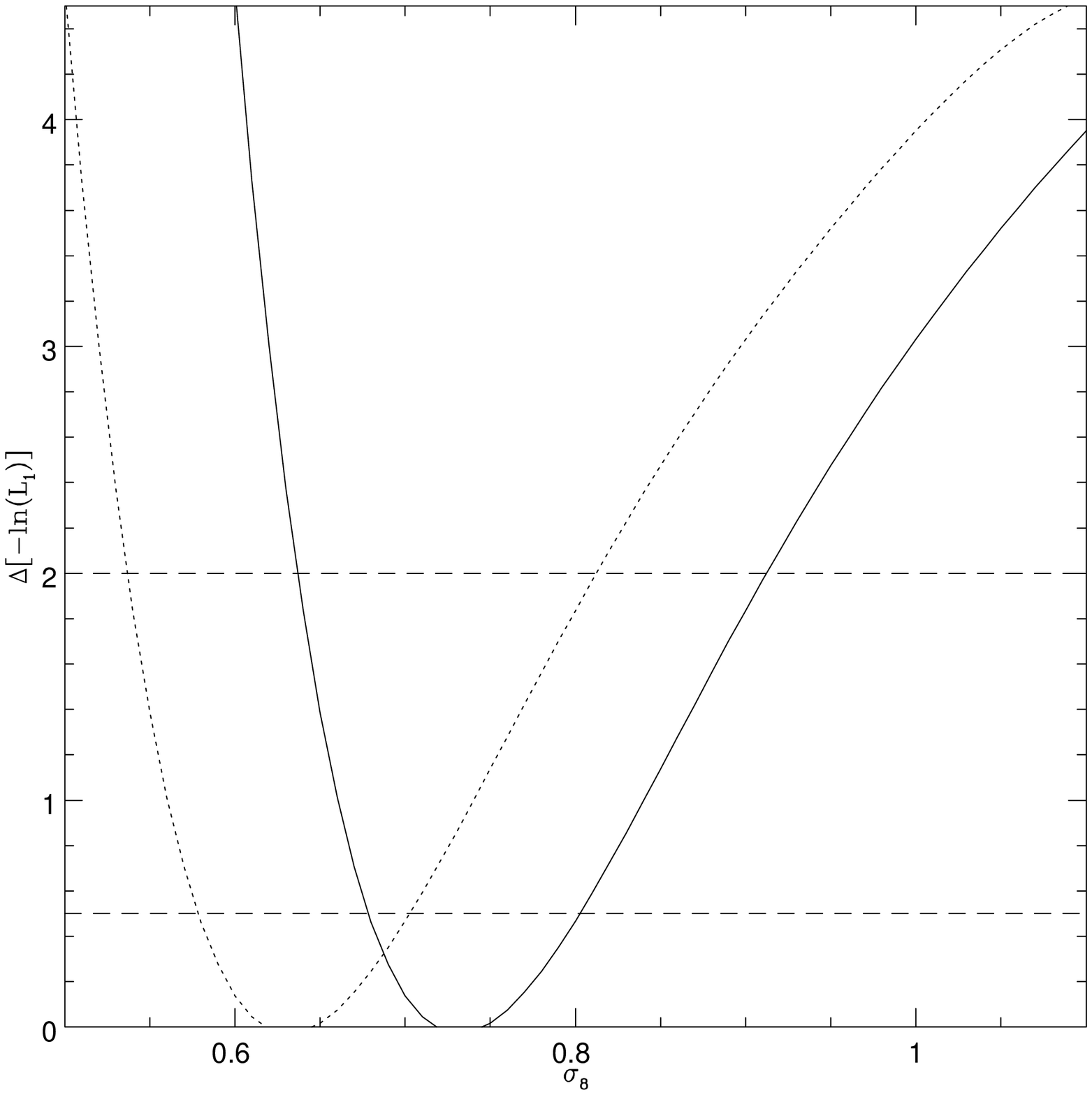,width=6cm}
}}
\caption{
$\Delta(-{\rm ln likelihood})$ for $\Omega_{\rm m}$ (left panel) and $\sigma_8$ (right panel), marginalized over the other two parameters (($\Gamma$, $\sigma_8$), left panel and ($\Omega_{\rm m}$, $\Gamma$), right panel). Solid lines are the prediction obtained from the model of this paper while the dotted lines those obtained from Eke et al. (1998).
Dashed lines are 1 and 2 $\sigma$ significance and the 3 $\sigma$ corresponds to the top of each panel.
}
\end{figure}

I get:


\begin{equation}
\frac{dN}{dT}=\frac{N_1(m,z)}{N_2(m,z)}
\end{equation}

\begin{eqnarray*}
N_1(m,z) &=&1/2\,{ m_1}\,\left (3+n\right )\rho\,\delta\,\sqrt {a}\sqrt
{2}A{m}^{{ m_2}+q}
\nonumber \\
& &
{e^{-\frac{1}{2C^2}\,{m}^{{ m_3}}a{\delta}^{2}-{a}^{1-
\alpha}{\delta}^{2-2\,\alpha}{m}^{{ m_3}}\left (C {m}^{{ m_4}}
\right )^{2\,\alpha}{ a_2}-1/2\,{a}^{1-2\,\alpha}{\delta}^{2-4\,
\alpha}{m}^{{ m_3}}\left (C {m}^{{ m_4}}\right )^{4\,\alpha}{{ a_2
}}^{2}}}
\nonumber \\
& &
\left (1+{ a_1}\,{a}^{-\alpha}{\delta}^{-2\,\alpha}\left ({C m
}^{{ m_4}}\right )^{2\,\alpha}\right )
\end{eqnarray*}
\begin{eqnarray*}
N_2(m,z) &=& {m}^{7/6}(3\,F{{ m_1}}^{2}{ LPh}\,q{m}^{{ a_3}}-3\,F{
m_1}\,{ a_3}\,{ m_{1l}}\,{ LPh}\,{m}^{{ a_3}}-2\,F{ LPh}\,{{
m_1}}^{2}{m}^{{ a_3}}
\nonumber \\
& &
+3\,F{{ m_1}}^{2}{ a_3}\,{ m_{1l}}\,{
LPh}\,{m}^{{ a_3}}
+2\,F{ LPh}\,{ m_1}\,{m}^{{ a_3}}
-3\,F{
m_1}\,{ LPh}\,q{m}^{{ a_3}}-3\,F{{ m_1}}^{2}{ a_3}\,{
LPh}\,{m}^{{ a_3}}
\nonumber \\
& &
+3\,F{ m_1}\,{ a_3}\,{ LPh}\,{m}^{{ a_3}}
-3\,F{{ m_1}}^{2}{ LPh}\,q-3\,F{ m_1}\,{ a_3}
+2\,F{ LPh}\,
{{ m_1}}^{2}-2\,{m}^{q}
-2\,{ t_1}\,{ m_1}\,{m}^{q}
\nonumber \\
& &
-3\,F{{ m_1}
}^{2}{ a_3}\,{ m_{1l}}\,{ LPh}+3\,F{ m_1}\,{ a_3}\,{ m_{1l}}
\,{ LPh}+3\,F{{ m_1}}^{2}{ a_3}-2\,F{ LPh}\,{ m_1}+3\,F{
m_1}\,{ LPh}\,q){ C}\,\sqrt {\pi}
\end{eqnarray*}

\begin{equation}
kT \simeq 8 keV m^{2/3}
\frac{
\left[
\frac{1}{m_1}+t_1^{\frac 23}
+\frac{K(m_1,x)}{(M_{15} m)^{8/3}}
\right]
}
{
\left[
\frac{1}{m_1}+t_1^{\frac 23}
 +\frac{K_0(m_1,x)}{M_{0}^{8/3}}
\right]
}
\end{equation}


%
%
%
where, I have defined:
$m=\frac{M}{M_8}$, 
$m_1=5/(n+3)$, 
$m_2=n/6$, $m_3=1+n/3$, $m_4=-1/2-n/6$, $a$, $a_1=0.06$, and $a_2=0.57$ 
$a_3=-5/(3m1)$, $m_{1l}=3m1/5+1$, $t_1=(t_{\Omega}/t)^{2/3}$, $r=2/3$, $q=8/3$, $\delta_{\rm c0}(z)=\delta$,
and $LerchPhi(m^{a3},1,m_{1l})=LPh$.

In order to use the same notation and variance of Henry (2000), the constant $C$ is defined as:
\begin{equation}
C=0.675\,\sqrt {{\frac {{\Gamma}(3+n)\sin(1/2\,n\pi )}{{2}^{n}n
\left (2+n\right )\left (1-n\right )\left (3-n\right )}}}\left (
 857.375\,{\frac {{k_{{o}}}^{3}{{\it Mpc}}^{3}}{{h}^{2}\Omega_{\rm m}}}\right
)^{-1/2-1/6\,n}
\end{equation}
and then $\sigma_8=\sigma(\Omega_{\rm m}, M=0.594 \times 10^{15} h^{-1} \Omega_{\rm m})$, and the XTF depends on the
parameters $n$, $k_{\rm o}$ and $\Omega_{\rm m}$.
The data that shall be fitted to the theory previously described, are those described in Section. 2 of Henry (2000).
I use a maximum likelihood fit to the unbinned data in order to determine various model parameters.
The method is described in Marshall et al (1983). Using Poisson probabilities, we have:
\begin{equation}
L_1 =\prod_i^N[\lambda (z_i,L_i)dzdLe^{-\lambda (z_i,L_i)dzdL}]\prod_je^{-\lambda (z_j,L_j)dzdL}
\end{equation}
where $\lambda (z,L)dzdL=\rho (z,L)\Omega (z,L)\frac{dV}{dz}dzdL$ is the expected number of objects in dzdL at z, L in the sample and the index j runs over all the differential elements in which no objects were observed. Defining $S=-2 ln(L_1)$ and dropping terms independent of the model parameters one finds:
%
%
\begin{equation}
S=-2\sum_{i=1}^{N}\ln \rho(z_{i},L_{i})
+2\int 
\int \rho(z,L)\Omega (z,L)\frac{d V(\Omega,z)}{dzd\Omega }dz dL
\label{eq:likelihh}
\end{equation}
Following (Henry 2000) notation and assumption Eq. (\ref{eq:likelihh}) is written as:
\begin{eqnarray}
S&=&-2\sum_{i=1}^{N}\ln \left[ n(\Omega _{m},z_{i},kT_{i})\frac{d^{2}V(\Omega _{m},z_{i})}{dzd\Omega }\right] +2\int_{kT_{\min }}^{kT_{\max }}dkT
\nonumber \\
& &
\int_{z_{\min }}^{z_{\max }}n(\Omega _{m},z,kT)\Omega (z,kT)\frac{d^{2}V(\Omega _{m},z)}{dzd\Omega }dz
\label{eq:likelih}
\end{eqnarray}
(Henry 2000),
where N is the number of clusters observed, $n(\Omega_{\rm m},z,kT)$ is the temperature function, $\Omega(z,kT)$ is the solid
angle in which a cluster with temperature kT at redshift z could have been detected (the selection
function) and $\frac{d^{2}V(\Omega _{m},z_{i})}{dzd\Omega }$ is the differential volume, which is given in the Appendix of Henry (2000) (see also Henry 2000 for a description of $\frac{d^{2}V(\Omega _{m},z_{i})}{dzd\Omega }$ in the first term of Eq. (\ref{eq:likelih})).
The best estimates for the parameter are obtained minimizing $S$.

At this point, we can fit the data described in Section. 2 of Henry (2000) to the theory previously described using
the quoted maximum likelihood method.
The most general description of the results requires the three parameters of the fit. I show these results
in Fig. 2, where I plotted the results for the open model.
It is straightforward to read off the value of n, which is
at the 68\% confidence for the open model.
These values shows that the correction introduced by the new form of the mass function and M-T relation gives rise to higher values of $\Omega_{\rm m}$ ($\Omega_{\rm m}=0.6 \pm 0.13$, while it is $\Omega_{\rm m}=0.49 \pm 0.12$ for Henry (2000)) and
$n=-1.5 \pm 0.32$ ($n=-1.72 \pm 0.34$ in Henry (2000)).
%
%
The presentation in Fig. 2a is somewhat difficult to appreciate,
so we also give the constraints for fewer parameters. In Fig. 2b,
I plot
$\Delta$(likelihood)
for the parameter $\Omega_{\rm m}$. The solid line is obtained
from the model of this paper while the dotted line is that
calculated by Henry (2000). The dashed lines represent various
confidence levels (68\%, 90\%, 95\%, 99\%). Constraints are
relatively tight when considering this single parameter. I find
that $\Omega_{\rm m}=0.6^{+0.12}_{-0.11}$ at the 68\% confidence
level and at $\Omega_{\rm m}=0.6^{+0.23}_{-0.2}$ the at 95\%
confidence level for the open model.

%
%
The constraints in Fig. 2a are plotted into a more conventional
format in Fig. 2c. Three parameters are still required, but the
constraints on n and $k_0$ are collapsed into $\sigma_8$.
Fig. 2c, plots the $68\%$ confidence contours for the parameters
$\sigma_8$, and $\Omega_{\rm m}$ for the open model (see also
Henry (2000), Fig. 9).
In Fig. 2d, I plot the constraints on $\Omega_{\Lambda}$ and
$\Omega_{\rm m}$ obtained using the same 25 clusters used in Henry
(2000), for the local sample, while the high redshift sample is
constituted from all the EMSS clusters with $z>0.3$ and RX
J0152.7-1357 (see Henry 2002). The solid lines are the 1 and 2
$\sigma$ contours obtained using the mass function and the M-T
relation of this paper, while the dashed line is the 1 $\sigma$
contour obtained using the PS mass function and the M-T relation
of Pierpaoli et al. 2001.

For a CDM spectrum, the expression for the XTF is much more complicated. It can be obtained combining Eq. (\ref{eq:nmm}),
Eq. (\ref{eq:temppp}), and our M-T relation.
The mass variance
can be obtained once a spectrum, $P(k)$, is fixed, by:
\begin{equation}
\sigma ^2(M)=\frac 1{2\pi ^2}\int_0^\infty dkk^2P(k)W^2(kR)
\label{eq:ma3}
\end{equation}
where $W(kR)$ is a top-hat smoothing function:
\begin{equation}
W(kR)=\frac 3{\left( kR\right) ^3}\left( \sin kR-kR\cos kR\right)
\label{eq:ma4}
\end{equation}
and the power spectrum $P(k)=Ak^nT^2(k)$ is fixed giving the transfer
function $T(k)$.
The CDM spectrum used in this paper is that of Bardeen et al. (1986)(equation~(G3)).
The shape of the spectrum and its amplitude at 8 $h^{-1}$ Mpc are described by 
$\sigma_8$ and $\Gamma$, respectively. Similarly to Eke et al. (1998)
the fitting parameters were $\Omega_{\rm m}$, $\Gamma$, $\sigma_8$, but just the results for 
$\Omega_{\rm m}$ and $\sigma_8$ were plotted.
%
%
Using the data used by Eke et al. (1998) (see their Sec. 3 \footnote{They combined the temperature data for 25 local clusters by Henry \& Arnaud (1991) with the sample of 10 {\it EMSS} clusters at $0.3<z<0.4$ by Henry et al. (1997)}) and the maximum likelihood parameter estimation given in Eke et al. (1998), Sec. 4,
%
%
%
%
%
I obtain the results plotted in Fig. 3a-b. Fig. 3a plots the
$\Delta(-{\rm ln likelihood})$ for $\Omega_{\rm m}$ (left panel)
and $\sigma_8$ (right panel), marginalized over the other two
parameters (($\Gamma$, $\sigma_8$) left panel, and ($\Omega_{\rm m}$, $\Gamma$), right panel). Solid lines are the prediction obtained from the model
of this paper while the dotted lines those obtained from Eke et
al. (1998). Dashed lines are 1 and 2 $\sigma$ significance and the
3 $\sigma$ corresponds to the top of each panel. Fig. 3 shows that
$\Omega_{\rm m}$ (left panel) and $\sigma_8$ (right panel) are
increased with respect to Eke et al. (1998) prediction: while in
Eke et al. (1998) $\Omega_{\rm m}=0.52^{+0.17}_{-0.16}$, and
$\sigma_8= 0.63^{+0.08}_{-0.05}$, I find that $\Omega_{\rm m}=
0.62^{+0.17}_{-0.15}$ and $\sigma_8= 0.73^{+0.07}_{-0.06}$. This
shows again an increase in $\Omega_{\rm m}$, also in qualitative
agreement with Eke et al. (1998) calculation taking account of
changes in the threshold for collapse suggested by Tozzi \&
Governato (1998).

\section{Results and discussion}

In this paper, I have revisited the constraints obtained
by several authors (Eke et al. 1998; Henry 2000)
on the estimated values of $\Omega_{\rm m}$, $n$ and $\sigma_8$ in the light of recent theoretical developments: new theoretical mass functions, a more accurate mass-temperature relation, also determined for arbitrary $\Omega_{\rm m}$ and $\Omega_{\rm \Lambda}$.
I repeated the Henry (2000) analysis but differently from the quoted paper, I changed the mass function and M-T relation, adopting again those of Del Popolo (2000a,b).
The new form of the mass function and M-T relation gives rise to higher values of $\Omega_{\rm m}$ ($\Omega_{\rm m}= 0.6 \pm 0.13$ in my estimation, while it is $\Omega_{\rm m}= 0.49 \pm 0.12$ for Henry (2000)) and
$n=-1.5 \pm 0.32$, in my estimation, while $n=-1.72 \pm 0.34$ in Henry (2000).
I have also obtained some constraints on $\Omega_{\Lambda}$ and $\Omega_{\rm m}$ obtained using the same 25 clusters used in Henry (2000), for the local sample, while the high redshift sample is constituted from all the EMSS clusters with $z>0.3$ and RX J0152.7-1357
(see Henry 2002). The 1 $\sigma$ contours obtained using the mass function and the M-T relation of this paper
shows that for $\Lambda=0$,
it is $0.32<\Omega_{\rm m}<0.57$ in the case of Henry (2002) and $0.43<\Omega_{\rm m}<0.73$ in my estimation. The figure shows the constraints to $\Omega_{\rm m}$ for different values of $\Omega_{\rm \Lambda}$.
Similar results to that obtained in the previous comparison with the Henry (2000) results
are obtained changing data and method, by following Eke et al. (1998). I obtain a value of $\Omega_{\rm m}= 0.62^{+0.17}_{-0.15}$, while in Eke et al. (1998) $\Omega_{\rm m}= 0.52^{+0.17}_{0.16}$, and $\sigma_8= 0.73^{+0.07}_{-0.06}$ while in Eke et al. (1998) $\sigma_8= 0.63^{+0.08}_{-0.05}$. This shows again an increase in $\Omega_{\rm m}$, also in agreement with Eke et al. (1998) calculation taking account of changes in the threshold for collapse suggested by Tozzi \& Governato (1998). 
\footnote{Replacing $\delta_{\rm c}$ with the $\delta_{\rm eff}$ of Tozzi \& Governato (1998), produces qualitatively similar changes on the mass function as those obtained using the Del Popolo (2002b) mass function.}
As previously told, this paper has its aim that of studying how ``systematic uncertainties" can influence the values of some cosmological parameters. It is well known that in literature the values obtained for $\Omega_{\rm m}$ span the range $ 0.2 \leq \Omega_{\rm m} \leq 1$ (Reichart et al. 1999).

Sadat et al. (1998) and Reichart et al. (1999)
Blanchard \& Bartlett (1998)
found results consistent with $\Omega_{\rm m}=1$. Viana \& Liddle (1999) found that
$\Omega_{\rm m}= 0.75$ with $\Omega_{\rm m}>0.3$ at the 90\% confidence level and $\Omega_{\rm m} \simeq 1$ still viable. Blanchard et al. (1998) found almost identical results ($\Omega_{\rm m} \simeq 0.74$ with $0.3<\Omega_{\rm m}<1.2$ at the 95\% confidence level).
Eke et al. (1998) found $\Omega_{\rm m}=0.45 \pm 0.2$. It is interesting to note (as previously mentioned) that Viana \& Liddle (1999) used the same data set as Eke et al. (1998) and showed that uncertainties both in fitting local data and in the theoretical modeling could significantly change the final results: they found $\Omega_{\rm m} \simeq 0.75$ as a preferred value with a critical density model acceptable at $<90\%$ c.l.

Different results were obtained by Bahcall et al. (1997) ($\Omega_{\rm m}=0.3 \pm 0.1$), Fan et al. (1997) ($\Omega_{\rm m}= 0.3 \pm 0.1$), Bahcall \& Fan (1998) ($\Omega_{\rm m}=0.2^{+0.3}_{-0.1}$) and several other authors.
More recent studies has been published, giving constraints on $\sigma_8$ and $\Omega_{\rm m}$: Pierpaoli et al. (2003) using a sample updated from HIFLUGCS they found a 68 \% confidence range of $\sigma_8=0.77^{+0.05}_{-0.04}$ for a standard $\Lambda$CDM model, while using the XLF from the REFLEX survey they obtain $\sigma_8=0.79^{+0.06}_{-0.07}$ for the same standard $\Lambda$CDM model. Schuecker et al. (2002), using the REFLEX cluster sample
and analyzing simultaneously the large scale clustering and the mean abundance of galaxy clusters get precise constraints on the normalized cosmic matter density $\Omega_{\rm m}$ ($\Omega_{\rm m}=0.341^{+0.031}_{-0.029}$) and the linear theory r.m.s. fluctuations in mass, $\sigma_8$ ($\sigma_8=0.711^+{0.039}_{-0.031}$). WMAP (Wilkinson Microwave Anisotropy Probing) precision data enabled accurate testing of cosmological models. Spergel et al. (2003) found that $\Omega_{\rm} h^2=0.135^{+0.008}_{-0.009}$, $\sigma_8=0.84 \pm 0.04$, where $h=0.71^{+0.04}_{-0.03}$. The formal random errors given 
in WMAP are of the 5 \% level while they are higher in our study. At this regard, I want to recall that the principal goal of the present study is not that of determining with high precision the values of the quoted cosmological parameters but to show how different values 
\footnote{As told in the introduction the results for $\Omega_{\rm m}$ obtained span the entire range of acceptable solutions: $0.2 \leq \Omega_{\rm m} \leq 1$}, 
discrepancies in the values of cosmological parameter quoted in literature  
are connected to a different mass function and M-T relation (the basic aim of the present work is to study how systematic
uncertainties in the application of the X-ray temperature function and
mass/X-ray temperature relation can influence the estimates of
important cosmological parameters).

The reasons leading to the quoted discrepancies has been studied
in several papers (Eke et al. 1998; Reichart et al. 1999; Donahue
\& Voit 1999; Borgani et al. 2001). According to Reichart (1999)
unknown systematic effects may be plaguing great part of the
quoted results.
Our analysis shows that improvements in the mass function and M-T relation increases the value of $\Omega_{\rm m}$. The effect of this increase is unable to enhance significantly the probability that $\Omega_{\rm m}=1$ in the case of
constraints like that of Fan, Bahcall \& Cen (1997) ($\Omega_{\rm m}=0.3$) or Bahcall \& Fan (1998) ($\Omega_{\rm m}=0.2$), and can give a small contribution even in the case of larger values for the value of the constraints obtained.
However, in any case it shows that even small correction in the physics of the collapse can induce noteworthy effects on the constraints obtained. Moreover, even changing the data or the way they are analyzed gives different results. As an example, changing their low-redshift sample, Donahue \& Voit (1999) showed a change in $\Omega_{\rm m}$ from 0.45 to 0.3.
Furthermore, as observations are reaching the first epoch of cluster assembly, treating them as dynamical relaxed and virialized systems is undoubtly an oversemplification. Hierarchical clustering scenario predicts that
a fraction between 0.3 and 0.6 of the $z=1$ population of clusters are observed less than 1 Gyr after the last major merger event and then are likely to be in a state of non-equilibrium. Although the quoted uncertainties has been so far of minor importance with respect to the paucity of observational data, a breakthrough is needed in the quality of the theoretical framework if high-redshift clusters are to take part in the high-precision-era of observational cosmology.

\begin{flushleft}

{\it Acknowledgements}

I am grateful to the referee P. Schuecker 
and to Prof. N. Ercan for helpful and stimulating
discussions during the period in which this work was performed. 
\end{flushleft}

\end{document}